\documentstyle{article}

\newcommand{\bea}{\begin{eqnarray}}
\newcommand{\eea}{\end{eqnarray}}

\begin{document}
\baselineskip = 13pt

\title{Cosmological Structure Problem of the Pre-Big Bang Scenario}
\author{Jai-chan Hwang \\
        Department of Astronomy and Atmospheric Sciences \\
        Kyungpook National University, Taegu, Korea}
\date{\today}
\maketitle
\parindent = 2em

\begin{abstract}

We calculate the density and gravitational wave spectrums generated in a 
version of string cosmology termed pre-big bang scenario.
The large scale structures are originated from quantum fluctuations
of the metric and dilaton field during a pole-like inflation stage
driven by a potential-less dilaton field realized in the low-energy
effective action of string theory.
The generated classical density field and the gravitational wave
in the second horizon crossing epoch show tilted spectrums with 
$n \simeq 4$ and $n_T \simeq 3$, respectively.
These differ from the observed spectrum of the large angular scale 
anisotropy of the cosmic microwave background radiation which supports 
scale invariant ones with $n \simeq 1$ and $n_T \simeq 0$.
This suggests that the pre-big bang stage is not suitable for generating
the present day observable large scale structures, and suggests the importance
of investigating the quantum generation processes during stringy era
with higher order quantum correction terms.

\end{abstract}



\section{Introduction}

String theory has attracted much attention in the literature
as a successful candidate for unifying gravity with other fundamental
interactions in the nature as a consistent theory of quantum gravity,
\cite{Superstring}.
However, effects of the string theory become significant on scales
far beyond the range of direct experimental test.
Theoretical successes provided by introducing an acceleration stage
in the early universe in solving the structure generation processes in
causal manner and other cosmological problems made the inflationary
idea settled as an orthodox paradigm in the cosmology community,
\cite{Inflation}.
{}From the observational side we have some important clues which 
provide constraints on theories of the early universe (thus, theories of the
high energy interactions) for the successful cosmology.
The most stringent constraint for filtering the viable theory
is provided by the large angular scale anisotropy of
the cosmic microwave background radiation (CMBR).
Both the amplitude and fluctuation spectrum of the microwave photons
can be used as the primary filters.

Combining above three pieces of ideas (string theories, inflation models, 
and cosmological observations) we can think of the following.
If we can make inflationary models based on string theories, the theories 
can be put under the testing ground of the cosmological observations.
There were some attempts to find viable inflationary models based on the 
low energy effective action of string theories. 
One successful candinate is the one introduced in \cite{pre-big-bang} termed 
``pre-big bang scenario'' which generically predicts a pole-like 
inflation stage, \cite{pre-big-bang}.
In this paper, we will calculate the generated density and gravitational wave 
spectrums and the consequent anisotropy in the CMBR, thus situating the theory 
under the screening process of cosmological observations.

Recently, we have presented a unified formulation of treating the
cosmological scalar type perturbations in generalized gravity theories,
\cite{GGT-H,GGT-HN,GGT-CT,GGT-QFT}.
Compared with previous work on the subject, a progress was
made by discovering a proper choice of the gauge which suits the problem.
The uniform-curvature gauge was introduced in \cite{H-PRW}
as one of several fundamental gauge choices available in
treating the cosmological perturbations.
{\it Classical evolution stage} is studied in \cite{GGT-H,GGT-HN,GGT-CT}:
Under the proper choice of the gauge (equivalently, gauge invariant variables)
the behavior in the large scale is characterized by a conserved quantity.
Ignoring the transient mode, the solution known in a minimally coupled
scalar field remains valid in the generalized gravity theories.
{\it Quantum generation stage} is studied in \cite{GGT-QFT}:
Using the uniform-curvature gauge, the quantum fluctuations generated
in generic phases of acceleration stage can be calculated in analytic forms.
The gravitational wave counterpart of the unified formulation (both for
handling the classical evolution and the quantum generation processes)
is recently presented in \cite{Kin-GW}.

In the point of view of our generalized gravity theories considered in 
\cite{GGT-H,GGT-QFT,Kin-GW} the low energy effective action of string theory 
can be regarded as a simple subset.
Thus, we will present the structure generation and evolution processes
in the string theory by reducing the general results in 
\cite{GGT-H,GGT-QFT,Kin-GW}.
Our study presented below will be based on the original frame of
the string theory.

\section{Cosmological setting for string theory}

The low-energy effective action of string theory is \cite{string-action}
\bea
   & & S = \int d^4 x \sqrt{-g} {1 \over 2} e^{-\phi}  
       \Big( R + \phi^{;a} \phi_{,a} \Big).
   \label{String-action}
\eea
The gravitational field equation and the equation of motion become
[eqs. (2,3) of \cite{GGT-H}]:
\bea
   R_{ab} = - \phi_{,a;b}, \quad
       \Box \phi = \phi^{;a} \phi_{,a}.
\eea

As a cosmological model describing the universe with structures,
we consider a homogeneous and isotropic (flat) background with
the general scalar and tensor type perturbations as
\bea
   d s^2 = - \left( 1 + 2 \alpha \right) d t^2 
       - \chi_{,\alpha} d t d x^\alpha
       + a^2 \left[ \delta_{\alpha\beta} \left( 1 + 2 \varphi \right)
       + 2 C^{(t)}_{\alpha\beta} \right] d x^\alpha d x^\beta,
   \label{metric-general}
\eea
where $\alpha ({\bf x}, t)$, $\chi ({\bf x}, t)$ and $\varphi ({\bf x}, t)$
are the scalar type metric perturbations;
$C^{(t)}_{\alpha\beta} ({\bf x}, t)$ is a tracefree-transverse
($C^{(t)\alpha}_{\;\;\;\;\;\alpha} = 0 = C^{(t)\beta}_{\;\;\;\;\;\alpha,\beta}$)
tensor type perturbation corresponding to the gravitational wave.
We have ignored the rotation mode in the metric whose behavior is trivially
characterized by the angular momentum conservation
[Sec. 3.2.2 of \cite{GGT1}]. 
We consider perturbations in the dilaton field as
$\phi ({\bf x}, t) = \bar \phi (t) + \delta \phi ({\bf x}, t)$
where a background quantity is indicated by an overbar
which will be neglected unless necessary.
Equations describing the background become [eqs. (20-22) of \cite{GGT-H}]:
\bea
   H^2 = H \dot \phi - {1\over 6} \dot \phi^2, \quad
       \ddot \phi + 3 H \dot \phi - \dot \phi^2 = 0,
   \label{BG}
\eea
where $H \equiv \dot a / a$.

\section{Formulation}
                             \label{sec:Formulation}

In our previous work \cite{GGT-H,GGT-HN,GGT-CT,GGT-QFT,Kin-GW}, we presented 
the classical evolution and the quantum generation processes (both for 
the scalar type perturbation and the gravitational wave) generally valid 
in a class of gravity theories represented by an action
\bea
   & & S = \int d^4 x \sqrt{-g} \left[ {1 \over 2} f (\phi, R)
       - {1\over 2} \omega (\phi) \phi^{;a} \phi_{,a} - V(\phi) \right].
   \label{GGT-action}
\eea
The string theory in eq. (\ref{String-action}) can be considered as 
a subset of theories in eq. (\ref{GGT-action}) with
$f = e^{-\phi} R$, $\omega  = - e^{-\phi}$, and $V = 0$.
In ths section we summarize the classical evolution and the quantum 
generation processes for handling the scalar type perturbation
and the gravitational wave which are quite {\it generally applicable} to
second-order gravity theories included in eq. (\ref{GGT-action}).

\subsection{Scalar type perturbation}

By introducing a gauge invariant combination 
\bea
   & & \delta \phi_\varphi
       \equiv \delta \phi - {\dot \phi \over H} \varphi
       \equiv - {\dot \phi \over H} \varphi_{\delta \phi},
   \label{UCG-UFG}
\eea
the second order perturbed action of the scalar type perturbation
of eq. (\ref{GGT-action}) becomes [eq. (29) of \cite{GGT-CT}]
\bea
   \delta^2 S = {1\over 2} \int a^3 Z \Bigg\{ \delta \dot \phi_\varphi^2
       - {1 \over a^2} \delta \phi_\varphi^{\;\; |\alpha}
       \delta \phi_{\varphi,\alpha}
       + {1 \over a^3 Z} {H \over \dot \phi}
       \left[ a^3 Z \left( {\dot \phi \over H} \right)^\cdot \right]^\cdot
       \delta \phi_\varphi^2 \Bigg\} dt d^3 x.
   \label{Action-pert}
\eea
The non-Einstein nature of the theory is present in a parameter $Z$ defined as
\bea
   & & Z (t) \equiv { \omega + {3 \dot F^2 \over 2 \dot \phi^2 F }
       \over \left( 1 + {\dot F \over 2 H F} \right)^2 },
   \label{Z-def}
\eea
where $F \equiv df / (dR)$.
The equations for the background in eq. (\ref{BG})
do not have Einstein limit.
However, for the perturbation part, by taking $Z = 1$ the perturbed action
in eq. (\ref{Action-pert}) reduces to the one valid in Einstein gravity
with a minimally coupled scalar field.
The equation of motion of the perturbed dilaton field follows as
\bea
   \delta \ddot \phi_\varphi
       + { (a^3 Z)^\cdot \over a^3 Z } \delta \dot \phi_\varphi
       - \left\{ {1 \over a^2} \nabla^2
       + {1\over a^3 Z} {H \over \dot \phi} \left[
       a^3 Z \left( {\dot \phi \over H} \right)^\cdot \right]^\cdot
       \right\} \delta \phi_\varphi = 0.
   \label{delta-phi-eq}
\eea
The large and small scale asymptotic solutions are, respectively:
\bea
   & & \delta \phi_\varphi ({\bf x}, t)
       = - {\dot \phi \over H} \varphi_{\delta \phi} ({\bf x}, t)
       = - {\dot \phi \over H} \left[ C ({\bf x}) - D ({\bf x}) \int^t
       {1 \over a^3 Z} {H^2 \over \dot \phi^2} dt \right],
   \label{delta-phi-LS-sol} \\
   & & \delta \phi_\varphi ({\bf k}, \eta)
       = { 1 \over a \sqrt{2k} } \Big[ c_1 ({\bf k}) e^{i k \eta}
       + c_2 ({\bf k}) e^{-ik\eta} \Big] {1 \over \sqrt{Z} },
   \label{delta-phi-SS-sol}
\eea
where $C({\bf x})$ and $D({\bf x})$ are integration constants of the
growing and decaying mode, respectively;
by relaxing the lower bound of the integration in $D({\bf x})$ term
we took into account of the possible mixing of the modes through
transition among different background evolution phases. 
$D({\bf x})$ term is higher order in the large scale expansion compared 
with the solutions in the other gauges; see \S VIA of \cite{GGT-HN}.
At this point $c_1({\bf k})$ and $c_2({\bf k})$ are arbitrary
integration constants.
We emphasize that the solutions in 
Eqs. (\ref{delta-phi-LS-sol},\ref{delta-phi-SS-sol}) are valid considering
general evolutions in $V(\phi)$, $\omega (\phi)$ and $f(\phi, R)$ as long as
the gravity theories belong to Eq. (\ref{GGT-action}) and second-order.
$C({\bf x})$, $D({\bf x})$, $c_1 ({\bf k})$ and $c_2 ({\bf k})$ are
integration constants considering such evolutions.

When we have $z^{\prime\prime} / z = n / \eta^2$ with $n = {\rm constant}$,
where $z (t) \equiv {a \dot \phi \over H} \sqrt{Z}$,
eq. (\ref{delta-phi-eq}) becomes a Bessel equation with a solution
[eq. (18) of \cite{GGT-QFT}]
\bea
   \delta \phi_{\varphi {\bf k}} (\eta)
       = {\sqrt{ \pi |\eta|} \over 2 a} \Big[ c_1 ({\bf k}) H_\nu^{(1)}
       (k|\eta|)
       + c_2 ({\bf k}) H_\nu^{(2)} (k|\eta|) \Big] {1 \over \sqrt{Z}}, \quad
       \nu \equiv \sqrt{ n + {1\over 4} }.
   \label{delta-phi-k-sol}
\eea
Considering $\delta \phi_{\varphi {\bf k}} (\eta)$ as a mode function
of $\delta \hat \phi ({\bf x}, t)$ which is regarded as a quantum 
Heisenberg operator, the canonical quantization condition leads to the 
following normalization condition \cite{GGT-QFT}
\bea
   & & | c_2 ({\bf k}) |^2 - | c_1 ({\bf k}) |^2 = 1.
   \label{normalization}
\eea
The quantization condition does not fix the mode function completely.
Choices of $c_1 ({\bf k})$ and $c_2 ({\bf k})$ depend on the vacuum state.
The positive frequency solution in the small scale limit corresponds to taking 
$c_2 = 1$ and $c_1 = 0$ which is the simplest choice; it is also the
choice {\it always} considered in the literature.
The power spectrum based on the vacuum expectation value becomes 
\bea
   & & {\cal P}_{\delta \hat \phi_\varphi} ( k, t )
       = {k^3 \over 2 \pi^2} \Big| \delta \phi_{\varphi {\bf k}} (t) \Big|^2.
   \label{Power-spectrum}
\eea

\subsection{Gravitational wave}
                                   \label{sec:GW}

The second order perturbed action for the gravitational wave part of 
eq. (\ref{GGT-action}) becomes \cite{Kin-GW}
\bea
   \delta^2 S_g = {1 \over 2} \int a^3 F
       \left( \dot C^{(t)\alpha}_{\;\;\;\;\beta} 
       \dot C^{(t)\beta}_{\;\;\;\;\alpha}
       - {1 \over a^2} C^{(t)\alpha}_{\;\;\;\;\beta,\gamma}
       C^{(t)\beta|\gamma}_{\;\;\;\;\alpha} \right) dt d^3 x.
   \label{GW-action}
\eea
The equation of motion is 
\bea
   \ddot C^{(t)}_{\alpha\beta}
       + \left( 3 H + {\dot F \over F} \right) \dot C^{(t)}_{\alpha\beta}
       - {1 \over a^2} \nabla^2 C^{(t)}_{\alpha\beta} = 0.
   \label{GW-eq-C}
\eea
The large and small scale asymptotic solutions are, respectively:
\bea
   & & C_{\alpha\beta}^{(t)} ({\bf x}, t)
       = C_{\alpha\beta} ({\bf x}) - D_{\alpha\beta} ({\bf x}) 
       \int^t {1\over a^3 F} dt,
   \label{GW-LS-sol} \\
   & & C_{\alpha\beta}^{(t)} ({\bf k}, \eta)
       = { 1 \over a \sqrt{F} } 
       \Big[ c_{1 \alpha\beta} ({\bf k}) e^{ik\eta} 
       + c_{2 \alpha\beta} ({\bf k}) e^{-ik\eta} \Big]. 
   \label{GW-SS-sol}
\eea
Similar comments below Eq. (\ref{delta-phi-SS-sol}) also apply to 
Eqs. (\ref{GW-LS-sol},\ref{GW-SS-sol}).
In order to handle the quantum generation process we consider a Hilbert space
operator $\hat C_{\alpha\beta}^{(t)}$ instead of the classical metric
perturbation $C^{(t)}_{\alpha\beta}$.
We write
\bea
   \hat C_{\alpha\beta}^{(t)} ({\bf x}, t)
       \equiv \int {d^3 k \over (2 \pi)^{3/2} } 
       \left[ \sum_\ell e^{i {\bf k} \cdot {\bf x}} \tilde h_{\ell {\bf k}} (t) 
       \hat a_{\ell {\bf k}} e_{\alpha\beta}^{(\ell)} ({\bf k}) 
       + {\rm h.c.} \right],
   \label{C-decomp-q}
\eea
where ${\ell} = +, \times$;
$e_{\alpha\beta}^{(\ell)}$ is bases of two ($+$ and$\times$) 
polarization states and $\hat a_{\ell}$ is the creation and annihilation 
operators of the polarization states.
By introducing
\bea
   \hat h_{\ell} ({\bf x}, t)
       \equiv \int {d^3 k \over (2 \pi)^{3/2} } 
       \Big[ e^{i {\bf k} \cdot {\bf x}}
       \tilde h_{\ell {\bf k}} (t) \hat a_{\ell {\bf k}} + {\rm h.c.} \Big],
   \label{h-def-q}
\eea
eq. (\ref{GW-action}) can be written as \cite{Ford-Parker}
\bea
   \delta^2 S_g 
       = \int a^3 F \sum_{\ell} \left( \dot {\hat h}_{\ell}^2
       - {1 \over a^2} \hat h_{\ell}^{\;\; |\gamma} 
       \hat h_{{\ell},\gamma} \right) dt d^3 x.
   \label{GW-action-h} 
\eea
{}For $z_g^{\prime\prime}/z_g = n_g/\eta^2$ with $n_g = {\rm constant}$, 
where $z_g (t) \equiv a \sqrt{F}$,
eq. (\ref{GW-eq-C}) has an exact solution.
In terms of the mode function we have
\bea
   \tilde h_{\ell {\bf k}} (\eta) 
       = {\sqrt{ \pi |\eta|} \over 2 a} 
       \Big[ c_{\ell 1} ({\bf k}) H_{\nu_g}^{(1)} (k|\eta|)
       + c_{\ell 2} ({\bf k}) H_{\nu_g}^{(2)} (k|\eta|) \Big] 
       \sqrt{ 1 \over 2 F }, 
   \label{GW-mode-sol}
\eea
where according to the canonical quantization condition
the coefficients $c_{\ell 1} ({\bf k})$ and $c_{\ell 2} ({\bf k})$ follow 
\bea
   \left| c_{\ell 2} ({\bf k}) \right|^2 
       - \left| c_{\ell 1} ({\bf k}) \right|^2 = 1.
   \label{c-normalization}
\eea
The simplest vacuum choice corresponds to taking
$c_{\ell 2} = 1$ and $c_{\ell 1} = 0$.
The power spectrum of the Hilbert space graviational wave 
operator based on the vacuum expectation value becomes
\bea
   {\cal P}_{\hat C^{(t)}_{\alpha\beta}} ({\bf k}, t)
       = 2 \sum_\ell {\cal P}_{\hat h_\ell} ({\bf k}, t)
       = 2 \sum_\ell {k^3 \over 2 \pi^2} 
       \big| \tilde h_{\ell {\bf k}} (t) \big|^2.
   \label{Power-q}
\eea
Previous studies on the evolution of the gravitational wave in 
some generalized gravity theories can be found in \cite{GW-GGT,GGT1,H-PRW}.

Using $\epsilon_i$'s introduced in eq. (87) of \cite{GGT-HN} we have
\bea
   {z_g^{\prime\prime} \over z_g} = a^2 H^2 \left( 1 + \epsilon_3 \right)
       \left( 2 + \epsilon_1 + \epsilon_3 \right)
       + a^2 H \dot \epsilon_3.
\eea
{}For $\dot \epsilon_i = 0$ we have
\bea
   n = { ( 1 - \epsilon_1 + \epsilon_2 - \epsilon_3 + \epsilon_4 )
       ( 2 + \epsilon_2 - \epsilon_3 + \epsilon_4 ) \over 
       ( 1 + \epsilon_1 )^2 }, \quad
       n_g = { ( 1 + \epsilon_3 ) ( 2 + \epsilon_1 + \epsilon_3 )
       \over (1 + \epsilon_1)^2 }.
   \label{n}
\eea

\section{Vacuum Fluctuations in the Pre-Big bang Scenario}

In the low energy effective action of the string theories, we have
\bea
   F = e^{-\phi}, \quad
       Z = {1 \over 2} e^{-\phi} \left( 1 - {\dot \phi \over 2H} \right)^{-2}.
\eea
{}From eq. (87) of \cite{GGT-HN} we have:
\bea
   \epsilon_1 \equiv {\dot H \over H^2}, \quad
       \epsilon_2 \equiv {\ddot \phi \over H \dot \phi}, \quad
       2 \epsilon_3 = \epsilon_4 = - {\dot \phi \over H}.
   \label{epsilons}
\eea
Equation (\ref{BG}) has the following solution
\bea
   & & a_\mp \propto | t - t_0 |^{\mp 1/\sqrt{3} }, \quad
       e^{\phi_\mp} \propto | t - t_0 |^{-1 \mp \sqrt{3} }.
   \label{BG-string}
\eea
The $a_{-}$ branch with $t < t_0$ represents a pole-like inflation stage
which is called a pre-big bang stage \cite{pre-big-bang}.
We will consider this case in the following.
We have
\bea
   \eta = - {3 - \sqrt{3} \over 2} {t_0 - t \over a}, \quad
       {\dot \phi \over H} = 3 + \sqrt{3}, \quad
       \delta \phi_\varphi = - ( 3 + \sqrt{3} ) \varphi_{\delta \phi}, \quad
       Z = ( 2 - \sqrt{3} ) e^{-\phi}.
   \label{UCG-UFG-string} 
\eea

{}For the scalar type perturbation
eqs. (\ref{Action-pert},\ref{delta-phi-eq},\ref{delta-phi-LS-sol}) become
\bea
   & & \delta^2 S = {1\over 2} \int a^3 Z \Bigg( \delta \dot \phi_\varphi^2
       - {1 \over a^2} \delta \phi_\varphi^{\;\; |\alpha}
       \delta \phi_{\varphi,\alpha} \Bigg) dt d^3 x,
   \\
   & & \delta \ddot \phi_\varphi - {1 \over t_0 - t} \delta \dot \phi_\varphi
       - {1 \over a^2} \nabla^2 \delta \phi_\varphi = 0,
   \label{delta-phi-eq-string} \\
   & & \delta \phi_\varphi ({\bf x}, t) = - ( 3 + \sqrt{3} ) 
       \Bigg[ C ({\bf x}) + {D ({\bf x}) \over 3 (3 + \sqrt{3})}
       \left( {|\eta| \over a^2 Z} \right)_1 \ln{(1 - t/t_0)} \Bigg].
   \label{delta-phi-LS-sol-string}
\eea
Since $ |\eta| / (a^2 Z) = {\rm constant}$ we evaluated it 
at an arbitrary epoch $t_1$.
$D({\bf x})$ term in Eq. (\ref{delta-phi-LS-sol-string}) logarithmically 
diverges near the end of the pre-big bang phase as $t$ approaches $t_0$.
Equation (\ref{epsilons}) becomes $\epsilon_1 = \epsilon_2 = \sqrt{3}$, 
$2 \epsilon_3 = \epsilon_4 = - 3 - \sqrt{3}$.
{}From eq. (\ref{n}) we have $n = -1/4$, thus $\nu = 0$.
Thus, eq. (\ref{delta-phi-k-sol}) becomes
\bea
   \delta \phi_{\varphi {\bf k}} (\eta)
       = { \sqrt{\pi} \over 2} \left( \sqrt{|\eta| \over a^2 Z} \right)_1 
       \Big[ c_1 ({\bf k}) H_0^{(1)} (k|\eta|)
       + c_2 ({\bf k}) H_0^{(2)} (k|\eta|) \Big].
   \label{delta-phi-k-sol-string}
\eea
The general power spectrum can be found from eqs. 
(\ref{Power-spectrum},\ref{delta-phi-k-sol-string}).
In the large scale limit we have
\bea
   {\cal P}^{1/2}_{\delta \hat \phi_\varphi} ({\bf k}, \eta)
   &=& (3 + \sqrt{3}) {\cal P}^{1/2}_{\hat \varphi_{\delta \phi}} 
       ({\bf k}, \eta)
   \nonumber \\
   &=& { 3 + \sqrt{3} \over \sqrt{6} }
       \left[ { H \over \sqrt{Z} } \left({k |\eta| \over \pi} \right)^{3/2} 
       \right]_1 \Big| \ln{(k|\eta|)} \Big|
       \times \Big| c_2 ({\bf k}) - c_1 ({\bf k}) \Big|,
   \label{P-QF}
\eea
where we used $\eta = - {\sqrt{3} - 1 \over 2} {1 \over a H}$.
Authors of \cite{Brustein-etal} derived the power spectrum of 
quantum fluctuations in a string era in a context of conformally related
Einstein frame assuming the simplest vacuum choice.

{}For the gravitational wave we can apply the formulation in 
\S \ref{sec:GW} with $F = e^{-\phi}$.
{}From eq. (\ref{n}) we have $n_g = - {1 \over 4}$, thus $\nu_g = 0$.
The general power spectrum follows from eqs. (\ref{GW-mode-sol},\ref{Power-q}). 
In the large scale limit we have
\bea
   {\cal P}^{1/2}_{ \hat C_{\alpha\beta}^{(t)} } ({\bf k}, \eta)
       = \sqrt{2} 
       \left[ { H \over \sqrt{Z} } \left({k |\eta| \over \pi} \right)^{3/2} 
       \right]_1 \Big| \ln{(k|\eta|)} \Big|
       \times \sqrt{ {1 \over 2} \sum_\ell \Big| c_{\ell 2} ({\bf k}) 
       - c_{\ell 1} ({\bf k}) \Big|^2 }.
   \label{P-GW-QF} 
\eea
The gravitational wave power spectrum in the pre-big bang scenario has been 
studied in \cite{PBB-GW,Brustein-etal} assuming the simplest vacuum choice.

By comparing eqs. (\ref{P-QF},\ref{P-GW-QF}), and taking the simplest vacuum 
choices, we have
\bea
   {\cal P}^{1/2}_{\hat C_{\alpha\beta}^{(t)}} ({\bf k}, t)
       = 2 \sqrt{3} \times
       {\cal P}^{1/2}_{\hat \varphi_{\delta \phi}} ({\bf k}, t).
\eea

\section{Generated Classical Spectrums}
                                 \label{sec:Classical}

In the large scale limit, ignoring the decaying mode, from 
eq. (\ref{delta-phi-LS-sol}) we have
\bea
   {\cal P}^{1/2}_C ({\bf k}) 
       = {\cal P}^{1/2}_{\varphi_{\delta \phi}} ({\bf k},t)
       \equiv {\cal P}^{1/2}_{\hat \varphi_{\delta \phi}} ({\bf k},t) 
       \times Q^{1/2} ({\bf k}),
   \label{C-ansatz}
\eea
where ${\cal P}_f$ and ${\cal P}_{\hat f}$ are power spectrums
based on the space averaging of the classically fluctuating field 
$f ({\bf x},t)$ and on the vacuum expectation value of the fluctuating 
quantum operator $\hat f ({\bf x}, t)$, respectively.
$Q(k)$ is a {\it classicalization factor} which may take into account of
possible effects from the classicalization processes \cite{Hu}.
The second step in eq. (\ref{C-ansatz}) should be considered as an ansatz.
Similarly, for the gravitational wave we take the following ansatz:
in the large scale limit during inflation era we assume
\bea
   {\cal P}_{C_{\alpha\beta}} ({\bf k})
       = {\cal P}_{C^{(t)}_{\alpha\beta}} ({\bf k},\eta)
       = 2 \sum_\ell {\cal P}_{h_\ell} ({\bf k},\eta)
       \equiv 2 \sum_\ell {\cal P}_{\hat h_\ell} ({\bf k}, \eta) \times 
       Q_\ell ({\bf k}),
   \label{GW-ansatz}
\eea
where $Q_\ell ({\bf k})$ is a classicalization factor for the 
gravitational wave with a polarization state $\ell$.

Thus, finally, from eqs. (\ref{C-ansatz},\ref{P-QF}) 
and eqs. (\ref{GW-ansatz},\ref{P-GW-QF}) we have 
\bea
   & & {\cal P}^{1/2}_{C} ({\bf k}) 
       = {\cal P}^{1/2}_{\varphi_{\delta \phi}} ({\bf k}, \eta) 
       = { 1 \over \sqrt{6} }
       \left[ { H \over \sqrt{Z} } \left({k |\eta| \over \pi} \right)^{3/2} 
       \right]_1 \Big| \ln{(k|\eta|)} \Big|
       \times \Big| c_2 ({\bf k}) - c_1 ({\bf k}) \Big| \sqrt{Q(k)},
   \label{P-C} \\
   & & {\cal P}^{1/2}_{ C_{\alpha\beta} } ({\bf k})
       = {\cal P}^{1/2}_{ C_{\alpha\beta}^{(t)} } ({\bf k}, \eta)
       = \sqrt{2} 
       \left[ { H \over \sqrt{Z} } \left({k |\eta| \over \pi} \right)^{3/2} 
       \right]_1 \Big| \ln{(k|\eta|)} \Big|
   \nonumber \\
   & & \qquad \qquad \qquad \qquad \qquad \qquad
       \times \sqrt{ {1 \over 2} \sum_\ell \Big| c_{\ell 2} ({\bf k}) 
       - c_{\ell 1} ({\bf k}) \Big|^2 \times Q_\ell ({\bf k}) },
   \label{P-GW} 
\eea
where the right hand sides should be evaluated while the scale
stays in the large scale during the pre-big bang era.
[The logarithmic time dependent terms in eqs. (\ref{P-C},\ref{P-GW}) 
look inconsistent because $C$ and $C_{\alpha\beta}$ are time independent.
We {\it guess} this point occurs because we have considered the quantum 
fluctuations in the full field $\varphi_{\delta \phi}$ contributing to the 
pure growing mode characterized by $C ({\bf x})$, and similarly for the 
gravitational wave.]
By comparing eqs. (\ref{P-C},\ref{P-GW}), taking the simplest vacuum 
choices and ignoring the classicalization factors, we have
\bea
   {\cal P}^{1/2}_{C^{(t)}_{\alpha\beta}} ({\bf k}, \eta)
       = 2 \sqrt{3} 
       \times {\cal P}^{1/2}_{\varphi_{\delta \phi}} ({\bf k}, \eta).
   \label{GW-scalar}
\eea

Up to this point we have calculated the quantum fluctuations which are
pushed outside the horizon, classicalized, and imprinted in the
spatial coeffiecients of the classical solutions in
eqs. (\ref{delta-phi-LS-sol},\ref{GW-LS-sol}).
These solutions characterize the large scale evolutions.
Now, we would like to argue that {\it the large scale evolution of the 
perturbations is characterized by conserved quantities independently of 
the changes in the background equation of state} 
(including the transition from the string dominated era to the ordinary 
radiation or matter dominated universe),
which can affect physics in the subhorizon scale.
The {\it scenario} we have in mind is the one in which the observationally 
relevant fluctuations become superhorizon scale (thus, large scale) 
during the acceleration stage and come back into the subhorizon scale during 
the ordinary matter dominated era effectively governed by Einstein's gravity.
{}From eqs. (\ref{delta-phi-LS-sol},\ref{GW-LS-sol}), ignoring the decaying 
modes which is higher order for $\varphi_{\delta \phi}$, we have
\bea
   \varphi_{\delta \phi} ({\bf x}, t) = C({\bf x}), \quad
       C^{(t)}_{\alpha\beta} ({\bf x}, t) = C_{\alpha\beta} ({\bf x}).
   \label{conservation}
\eea
Thus, in large scale limit the growing modes of $\varphi_{\delta \phi}$
and $C^{(t)}_{\alpha\beta}$ are conserved.
We note that this conserved behavior is generally valid for 
a class of generalized gravity which includes our string theory;
i.e., considering general $V(\phi)$, $\omega(\phi)$, and $f(\phi,R)$ in
eq. (\ref{GGT-action}). 
Let us make this point more clear:
As long as the two conditions [first, the transition process involves 
only the gravity theories which are subsets of Eq. (\ref{GGT-action}), and 
second, the scale remains in the large scale during the transition] are held, 
we have a general large scale solution in Eq. (\ref{delta-phi-LS-sol}) which is
\bea
   & & \varphi_{\delta \phi} ({\bf x}, t)
       = C ({\bf x}) 
       - D ({\bf x}) \int^t {1 \over a^3 Z} {H^2 \over \dot \phi^2} dt.
   \label{varphi-conserv}
\eea
Remarkably, this equation (and Eq. [\ref{GW-LS-sol}]) is valid considering 
general evolutions of $V(\phi)$, $f(\phi,R)$, $\omega(\phi)$ etc., 
as long as the evolution occurs satisfying the two conditions.
As mentioned before, $C({\bf x})$ and $D({\bf x})$ are ``integration constants''
considering the general evolutions.
Thus, as long as the two conditions are met, Eq. (\ref{varphi-conserv}) 
takes care of the general evolution from the pre-big bang phase to 
Einstein gravity, and although there can arise the
mixing in the $D({\bf x})$ mode, the growing $C({\bf x})$ will not be affected.
This argument justify ignoring the decaying mode in the large scale evolution; 
the growing $C({\bf x})$ mode in Einstein phase is only affected by the 
same $C({\bf x})$ mode in the pre-big bang phase.  

This conservation in the large scale may reflect the kinematic nature 
of the evolution in the superhorizon scale.
Equation (\ref{conservation}) is valid in a scale larger than the sound
horizon for the scalar type perturbation and the visual horizon for the
gravitational wave.
Notice that the generalized nature of the gravity theory, $Z$ or $F$, does not
appear in eq. (\ref{conservation}).
In Einstein's gravity with the ordinary fluid and the general equation of state,
i.e., $p = p(\mu)$, we have the same conserved behavior again as
in eq. (\ref{conservation}), but now $\varphi_{\delta \phi}$ being replaced
by $\varphi_\Psi$, where $\Psi = 0$ corresponds to taking the 
comoving gauge, \cite{H-PRW}.
[In Einstein gravity the uniform-field gauge is the same as the 
comoving gauge with $\Psi = - \dot \phi \delta \phi$, \cite{Bardeen}.
{}For thorough discussions concerning the large scale conserved quantities 
in the fluid era and the minimally coupled scalar field, see \cite{conserved}.]

Armed with this knowledge from previous studies, we can understand the
classical evolution of the perturbations in the large scale in the following
manner:
In addition to the quantum fluctuation of the tranverse-tracefree
part of the metric, the quantum fluctuation in the dilaton field simultaneously 
excites and accompanies quantum fluctuations in the scalar type metric.
During the acceleration era the relevant scale becomes superhorizon size
and becomes classical.
While the perturbations are superhorizon size they can be characterized 
by the conserved quantities $C({\bf x})$ and $C_{\alpha\beta} ({\bf x})$ 
which are in fact $\varphi_{\delta \phi}$ and $C^{(t)}_{\alpha\beta}$, 
respectively. 
[For handling the gravitational wave the often favored method in the literature
is matching the Bogoliubov coefficients assuming sudden jump transitions
among different background expansion stages \cite{Bogoliubov}.
In this paper we use the general conservation property of the gravitational 
wave in the large scale; this method was originally used in \cite{Starobinsky}.]
We do not imply that everything is conserved, but rather imply that, 
from the conserved quantity the behavior of every other variable follows
as linear combination.
In the linear theory, all variables are linearly related with each other.
{}From any one known solution we can derive all the other.
Thus, we can regard the informations of the spatial structures are encoded 
in $C({\bf x})$ and $C_{\alpha\beta} ({\bf x})$, and this informations are 
preserved as long as the scale remains in the large scale 
(Jeans scale in the fluid era for the scalar type perturbation).
In this sense, changing phases of the underlying background universe
including the gravity sector and the equation of state do not affect
the evolution of structures in the superhorizon scale.
[This is true, at least, {\it as long as} the gravity is included in 
eq. (\ref{GGT-action}).
We expect (but yet to be proved!) it to be more generally true reflecting 
the kinematic nature of physics in the superhorizon scale.]
Thus, as long as the graceful exit problem in the pre-big bang scenario can 
be settled with reasonable modifications of the gravity sector, our description 
of the superhorizon size perturbations in terms of a conserved quantity is 
expected to remain valid; for the graceful exit problem in the pre-big bang 
scenario, see \cite{graceful-exit,Rey}.
As the scale comes back inside horizon, informations about observationally
relevant quantities can be decoded from the conserved quantities $C({\bf x})$
and $C_{\alpha\beta} ({\bf x})$.

In the matter dominated era, the solution for the relative density
fluctuation ($\delta \equiv {\delta \rho / \rho}$) becomes
[eq. (23) of \cite{GGT-H}]
\bea
   & & \delta ({\bf x}, t) = {2 \over 5} \left( {k \over a H} \right)^2 
       C ({\bf x}).
   \label{delta-MDE}
\eea
Thus, in the second horizon crossing, where ${k \over aH} |_{\rm HC} \equiv 1$,
we have 
\bea
   {\cal P}^{1/2}_{\delta} (k,t_{\rm HC})
   = {2 \over 5} {\cal P}^{1/2}_{C} (k)
   = { (a H)^2 \over \sqrt{2} \pi } k^{-1/2} |\delta_{\bf k} (t)|.
   \label{HC-power-spectrum}
\eea
Conventionally we take $|\delta_{\bf k} (t)|^2 \equiv A(t) k^n$
where $n$ is a spectral index; $n = 1$ corresponds to the Zeldovich spectrum.
Ignoring the classicalization factor ($Q \equiv 1$),
the vacuum dependence ($c_2 \equiv 1$ and $c_1 \equiv 0$), and
the logarithmic dependence on $k$, 
from eqs. (\ref{P-C},\ref{HC-power-spectrum}) we have
a tilted spectrum with $n = 4$.

Similarly for the gravitational wave, in the matter dominated era and 
ignoring the transient mode, we conventionally take
\bea
   {\cal P}_{C^{(t)}_{\alpha\beta}}^{1/2} ({\bf k}, t)
       = {\sqrt{2} \over \pi} A_T^{1/2} ({\bf k}) 
       \Bigg| {3 j_1 (k\eta) \over k \eta} \Bigg|.
   \label{P-GW-MDE}
\eea
By matching eq. (\ref{P-GW}) with eq. (\ref{P-GW-MDE}) in the large scale limit
we can show that 
\bea
   A_T^{1/2} ({\bf k}) = \pi 
       \left[ { H \over \sqrt{Z} } \left({k |\eta| \over \pi} \right)^{3/2} 
       \right]_1 \Big| \ln{(k|\eta|)} \Big|
       \times \sqrt{ {1 \over 2} \sum_\ell \Big| c_{\ell 2} ({\bf k})
       - c_{\ell 1} ({\bf k}) \Big|^2 \times Q_\ell ({\bf k}) }.
   \label{A_T}
\eea
One often writes $A_T ({\bf k}) \equiv A_T k^{n_T}$.
If we ignore the dependences on the choice of the vacuum state and the 
classicalization factor, we have $n_T = 3$.
This spectrum also differs from the scale invariant one with $n_T = 0$.
The gravitational waves generated in the pre-big bang scenarios
are thoroughly investigated in \cite{PBB-GW,Brustein-etal}.

The CMBR last scattered in the recombination era must have gone through 
the perturbed spacetime.
The spatially and temporally fluctuating spacetime metric causes the
directional (${\bf e}$) dependence of the CMBR temperature 
observed in a given position ${\bf x}_R$ as $\delta T({\bf e}; {\bf x}_R)$,
\cite{Sachs-Wolfe}.
Conventionally we expand 
${\delta T \over T} ({\bf e}; {\bf x}_R) 
= \sum_{lm} a_{lm} ({\bf x}_R) Y_{lm} ({\bf e})$.
We are often interested in the rotationally symmetric quantity 
$\langle a_l^2 \rangle \equiv 
\langle | a_{lm} ({\bf x}_R) |^2 \rangle_{{\bf x}_R}$ 
where $\langle \rangle_{{\bf x}_R}$ is an average over possible locations of 
an observer.
In the following we present the result in order to show the 
proper normalization with our notation. 
{}For the scalar type perturbation and the gravitational wave, we have,
respectively \cite{a_l}
\bea
   & & \langle a_l^2 \rangle
       = {4 \pi \over 25} \int_0^\infty 
       {\cal P}_C (k) j_l^2 (kx) d \ln{k}, \quad
       \langle a_l^2 \rangle
       = {9 \pi \over 2} {\Gamma(l + 3) \over \Gamma(l -1)}
       \int_0^\infty A_T (k) |I_l (k)|^2 d \ln{k}, 
   \nonumber \\
   & & I_l (k) \equiv {2 \over \pi} \int_{\eta_e}^{\eta_o}
       {j_2 (k \eta) \over k \eta}
       {j_l (k \eta_0 - k \eta) \over (k \eta_0 - k \eta )^2} k d \eta,
   \label{a_l}
\eea
where $x = 2/H_0$.
By integrating eq. (\ref{a_l}) using eqs. (\ref{P-C},\ref{A_T}) 
we can estimate the effects on the CMBR.  
The observational results favor $n \simeq 1$ and $n_T \simeq 0$, \cite{COBE}.

\section{Comparison with previous works and Discussions}

Studies of the scalar type fluctuations in pre-big bang scenario were
also carried out in \cite{Gasperini-Veneziano,Brustein-etal}.
The study in \cite{Gasperini-Veneziano} is based on a zero-shear gauge
which complicates the analysis compared with the
one based on the uniform-curvature gauge used in our approach.
Authors in \cite{Brustein-etal} introduced an off-diagonal gauge
which is in fact the same as the uniform-curvature gauge. 
Studies in \cite{Gasperini-Veneziano,Brustein-etal} are
made in a conformally related {\it Einstein frame}.
In the classical perturbation level the effective string theory 
in eq. (\ref{String-action}) and Einstein's gravity can be transformed 
to each other through the conformal rescaling; this is shown in more 
general context of eq. (\ref{GGT-action}) in \cite{GGT-CT}.
The equivalence in the fully quantum level is not obvious, \cite{Duff}.
However, in the quantum level of our perturbative semiclassical approximation, 
treating both the perturbed dilaton field and the perturbed metric 
as quantum operators, we can show the equivalence through the 
conformal transformation, see \cite{GGT-QFT}.
Our study presented above is directly made in the {\it original frame of
the string theory} both for the quantum generation and the classical evolution;
we may stress that it is hardly more difficult than working in Einstein frame.
Authors of \cite{Brustein-etal} derived only the power spectrum of the 
quantum fluctuations in the string era which corresponds to eq. (\ref{P-QF}).
We have been able to derive the final density spectrum at the second 
horizon epoch and present day universe (thus partly complete the structure 
formation) by {\it assuming} a transition of the underlying gravity, 
from the string theory to the Einstein one, while the observationally 
relevant perturbations were on superhorizon scale. 
The transition of the string stage into Einstein one is a serious assumption
which needs further investigation; 
though, we made an argument that as long as the perturbations we are 
interested in are in the superhorizon scale and as long as the gravity
belongs to Eq. (\ref{GGT-action}), we can apply the conservation 
argument of the perturbed curvature variable in the uniform field 
(or comoving) gauge as in eq. (\ref{conservation}); 
see below Eq. (\ref{GW-scalar}).
Equations (\ref{delta-MDE}-\ref{a_l}) are valid if the present day universe 
is governed by the Einstein gravity.

In this paper we have not estimated the realistic amplitude of the
observationally relevant quantities which may depend on the specific 
realization of the pre-big bang scenario. 
However, the power spectrums with $n \simeq 4$ and $n_T \simeq 3$ in 
eqs. (\ref{P-C},\ref{P-GW}) are in contrast with the observationally 
supported spectrum with $n \simeq 1$ and $n_T \simeq 0$, \cite{COBE}.
This implies that the spectrum generated in the above theory 
{\it cannot} be considered as the source for the observed fluctuations in the 
CMBR and also the observed large scale structures.
At this point, we would like to point out the possibility that the choices of 
the vacuum states and the classicalization factors in 
eqs. (\ref{P-C},\ref{P-GW}) can completely dominate the whole power spectrum.
Effects of these two factors cannot be estimated within the scope of our 
linear treatment of the quantum fluctuations, and may require other 
physical arguments.

In \S \ref{sec:Classical} we have derived the final observational results
based on a couple of important assumptions: 
The first assumption is that the observationally relevant scale exits 
the horizon to become a superhorizon size, and becomes classical during 
the pre-big bang era.
The second assumption is that at some epoch in the early universe
the gravity transits from the low energy string theory to Einstein theory,
and at the time of the transition the relevant perturbations were
superhorizon size.
Most of the observationally relevant scale may exit the horizon during the
last moments of the latest inflation stage. 
Thus, for the validity of the first assumption it is essential that 
the pre-big bang stage serves during near end of the latest inflation.

Recently, an interesting suggestion was made that the graceful exit problem 
in the original pre-big bang scenario can be resolved by including the
quantum back reaction effect \cite{Rey}.
Since the spacetime curvature term diverges near the pole region 
($t \simeq t_0$), the quantum back reaction could become important 
as the inflation approaches the pole.
In such a case, since the observationally relevant perturbations for the 
large scale structure leave the horizon most probably during the 
stringy pre-big bang era, one has to investigate the scalar and 
the gravitational wave spectrums based on an action which includes
higher order string quantum correction terms.
Results for observable quantities could be very different, and at the moment, 
the analysis is beyond the scope of our formulation used in this paper;
the pre-big bang scenario considered in this paper does not include 
effects of the dilaton potential, axion field, and quantum stringy effects 
like higher order curvature correction terms except for the $f(\phi, R)$ term.
[In this case, there remains a possibility that the $n \simeq 4$ and 
$n_T \simeq 3$ power spectrums could be interpreted as the spectrums of 
the present day superhorizon scale structures which exited the horizon 
before the stringy quantum era begins.]
The general formulation presented in \cite{GGT-H,GGT-HN,GGT-CT,GGT-QFT,Kin-GW} 
and summarized in \S \ref{sec:Formulation} will be a useful step for 
extended studies.
As a conclusion, although the pre-big bang inflation fails observationally,
one cannot blame the string theory on the charge of the faileur.
Investigating the structure generation processses in the strong quantum regime 
is a very important and interesting {\it open problem} which is left for 
future endeavour.

\subsection*{Acknowledgments}

We wish to thank Drs. R. Brustein, M. Gasperini, H. Noh and S.-J. Rey
for useful discussions and comments on the manuscript.
This work was supported by the Korea Science and Engineering Foundation, 
Grant No. 95-0702-04-01-3 and through the SRC program of SNU-CTP.


\end{document}